\documentstyle[epsf,12pt]{article}
\newcommand{\lsim}{\mbox{\raisebox{-.3em}{$\stackrel{<}{\sim}$}}}

\renewcommand{\cite}[1]{\ref{#1}}

\newcommand{\half}{\frac{1}{2}}

\newcommand{\beq}{\begin{equation}}
\newcommand{\eeq}{\end{equation}}
\newcommand{\beqa}{\begin{eqnarray}}
\newcommand{\eeqa}{\end{eqnarray}}
\newcommand{\bpr}{\begin{problem}}
\newcommand{\epr}{\end{problem}}
\newcommand{\bcent}{\begin{center}}
\newcommand{\ecent}{\end{center}}
\newcommand{\bfig}{\begin{figure}}
\newcommand{\efig}{\end{figure}}
\newcommand{\bpc}{\begin{picture}}
\newcommand{\epc}{\end{picture}}

\newcommand{\Dslash}{D\hspace{-.65em}/}

\newcommand{\nnb}{\nonumber}
\newcommand{\reflef}{(\ref}
\newcommand{\MP}{M_{\rm P}}
\newcommand{\psibar}{\overline{\psi}}

\begin{document}
\baselineskip=0.6cm

\bcent
{\Large\bf Possible link between the changing fine-structure constant and the accelerating universe via scalar-tensor theory
}\footnote{Delivered at First International ASTROD School and Symposium, September 12-23, 2001, Beijing.}\\[.5em]

 Yasunori Fujii\\

Nihon Fukushi University\\
Handa, Aichi, 475-0012 Japan

\ecent
\bigskip
\hspace*{1cm}
\begin{minipage}{12.5cm}
\bcent
{\large\bf Abstract}
\ecent
\baselineskip=.4cm
{\small
In 1976, Shlyakhter showed that the Sm data from Oklo results in the
upper bound on the time-variability of the fine-structure constant: $|\dot{\alpha}/\alpha| \lsim 
 10^{-17}{\rm y}^{-1}$, which has ever been the most stringent bound.
 Since the details have never been published, however, we recently
re-analyzed the latest data according to Shlyakhter's recipe.  We
nearly re-confirmed his result.  To be more precise, however, the Sm data gives
either an upper-bound or an ``evidence" for a changing $\alpha$:
$\dot{\alpha}/\alpha = -(0.44 \pm 0.04)\times 10^{-16}{\rm y}^{-1}$.  A
 remark is made  to a similar re-analysis due to Damour and Dyson.
We also compare our result with a recent ``evidence" due to Webb et
al, obtained from distant QSO's.  We point out a possible connection
between this time-dependence and the behavior of a scalar field
supposed to be responsible for the acceleration of the universe, also
revealed recently. 
}
\end{minipage}

\section{How did constants cease to be constant?}	

More than 60 years ago, Dirac startled the science community by saying
that the gravitational constant $G$ is not a true constant, but decays
like the inverse of the cosmic time $t$:$^1$
\begin{equation}
G(t) \sim t^{-1},
\label{bjn-1}
\end{equation}
from which follows that the rate of the fractional change $\dot{G}/G$ should be like
\beq
\frac{\dot{G}}{G} \sim -t^{-1} \sim -10^{-10}{\rm y}^{-1},
\label{bjn-2}
\eeq
because the present age of the universe is believed to be somewhere around $10^{10}{\rm y}$.

I do not know how many physicists were really convinced with Dirac's
argument, based on the ``Large-numbers Hypothesis."  No matter what
philosophy he had started with, however, it is clear that he opened up
a new view of looking at the very fundamental aspect of laws of
Nature.  Some people tried hard to prove or disprove his result.
But the number is so small that almost any noise would easily wipe out
the real effect if there is any.

So far no solid evidence for the changing $G$ has been reported.  We
have heard only about upper bounds, like 
\[
\frac{\dot{G}}{G} = \left\{
\begin{array}{l}
(0.2\pm 0.4)\times 10^{-11}{\rm y}^{-1},
\hspace{2em}\quad \mbox{\hspace{1em}Viking Project(1983),} \\[.5em]
(-0.06\pm 0.2)\times 10^{-11}{\rm y}^{-1},
\hspace{2em}\quad \mbox{Binary Pulsar}(1996). 
\end{array}
\right.
\]
These are already below Dirac's prediction.  The search for the
changing $G$ may appear to have lost the motivation.  We have now,
however, ``unified theories,`` attempts of unifying physics of gravitation and of microscopic world.   According to the theoretical models, including string theory as one of the best known examples, time-dependence of various coupling constants should be expected generically for the following reasons:

\begin{itemize}
\item Truly constant coupling constants in the theories at the deeper level in higher-dimensional spacetime.
\item At another level we have a realistic theory in 4 dimensions,
where the {\em effective} coupling constants to be measured are derived from those in the fundamental theories through the process of compactification.
\item In this process, some of the scalar fields, notably a dilaton, comes into play. The scalar fields are expected to change slowly as the universe evolves, and so do the effective coupling constants.
\end{itemize}

In this context, the way the coupling constants change may be different from the way Dirac envisioned.  Obviously coupling constants are not only confined to the gravitational coupling constant, but also include other coupling constants, notably the fine-structure constant $\alpha = e^2/(4\pi \hbar c)$, the square of the electromagnetic coupling constant and its strong interaction analogue $\alpha_s$.  The result of the past investigations on their time variability is summarized:
\begin{table}[h]
\bcent
\begin{tabular}{||l|l||l|l||} \hline\hline
\hspace{1em}\raisebox{-.5em}{Sources} & Look-back  & \hspace{.0em}\raisebox{-.5em}{$\dot{\alpha}/\alpha ({\rm y}^{-1})$}  &\hspace{0.0 em} \raisebox{-.5em}{$\dot{\alpha_s}/\alpha_s({\rm y}^{-1})$} \\[-.7em]
&time$({\rm y})$ && \\
 \hline
Primord nucl synth &&& \\[-.8em]
\hspace{1em}{\footnotesize Hoyle et al, 1965} &    \raisebox{.5em}{$1\times 10^{10}$ } & &  \raisebox{.5em}{$1\times 10^{-13}$} \\[-.2em]
Very long-lived nuclei &&& \\[-.8em]
\hspace{1em}{\footnotesize Dyson, 1967} & \raisebox{.5em}{$5\times 10^9$}  & \raisebox{.5em}{$3\times 10^{-13}$} &  \\[-.2em]
Stellar nuc synthesis &&& \\[-.8em]
\hspace{1em}{\footnotesize Davies, 1972} & \raisebox{.5em}{$5\times 10^9$} & &\raisebox{.5em}{$2\times 10^{-12}$}\\[-.2em]
Oklo phenomenon &&& \\[-.8em]
\hspace{1em}{\footnotesize Shlyakhter, 1976}& \raisebox{.5em}{$2\times 10^9$} & \raisebox{.5em}{$1\times 10^{-17} $} & \raisebox{.5em}{$ 5 \times 10^{-19}$ }\\ [-.2em]
Time standards&&& \\[-.8em]
\hspace{1em}{\footnotesize  Prestage  et al, 1995}   & \raisebox{.5em}{0} & \raisebox{.5em}{$3\times 10^{-13}$}  &\\ [-.2em]
Distant QSO &&& \\[-.8em]
\hspace{1em}{\footnotesize Webb et al, 2001} & \raisebox{.5em}{$1\times10^{10}$} &  \raisebox{.5em}{$7\times 10^{-16}$}  & \\ \hline\hline
\end{tabular}
\caption{\footnotesize Results for time-variability of $\alpha$ and $\alpha_s$.}
\ecent
\end{table}

Most of them are upper bounds, except for the last entry on the recent 
result due to Webb {\it et al.},  who claim the first ``evidence" of
the changing $\alpha$.$^2$  As another point to be noticed in this Table, we find that the ``Oklo phenomenon" gives an upper bound several orders of magnitude better than any other results.

\section{Oklo phenomenon}

Oklo is the name of a uranium mine in Gabon, West Africa, but should
be remembered as a place where the evidence of ``natural reactors" was
discovered and verified for the first time in 1974.$^3$  By natural
reactors, we mean that  self-sustained nuclear fission reactions
occurred {\em naturally} about 2 billion years ago, lasting some
million years.  This was a rediscovery of the theoretical prediction
due to Kuroda in 1955.$^4$

Though it might sound rather strange, this phenomenon took place for several reasons.  The most crucial among them is the fact that the abundance of $^{235}{\rm U}$ must have been  $\sim 3\%$, much higher than the present value $ 0.7207\%$ due to the difference between lifetimes of $^{235}{\rm U}$ and $^{238}{\rm U}$.  Since 1974, extensive studies have been made on the remnants of fission products in the so-called reactor zones in Oklo.

In 1976, Shlyakhter pointed out that observation of the isotope
$^{149}{\rm Sm}$ could be useful in determining how much different the
nuclear reactions 2 billion years ago were from what they are 
today.$^5$  The natural abundance of this isotope is 13.8\%, while the observation
in Oklo showed much smaller values.  This deficit can be understood in 
terms of the consumption due to the neutron absorption reaction:
\beq
n+^{149}\hspace{-.2em}{\rm Sm} \rightarrow ^{150}\hspace{-.2em}{\rm Sm} + \gamma,
\label{bjn-4}
\eeq
that had occurred inside natural reactors.

This scenario was verified by using the data obtained from today's
laboratories.  This implied that the parameters of the reaction
\reflef{bjn-4}) should not have been very much different from what
they are today.  This conclusion was made even more precise  by noting that this reaction is dominated by a resonance lying as low as $E_r =97.3 {\rm meV}$, which is smaller by more than 7 orders of magnitude than a typical energy scale $\sim 1{\rm MeV}$ in most of the nuclear reactions.  The half-width is also very narrow; 61.0meV.

In view of the smallness of $E_r$, we should consider that the
presence of this resonance is a consequence of a nearly complete
cancellation between two interactions; repulsive Coulomb energy and
the nuclear force due to the strong interaction, leaving a tiny
residual effect.  Suppose we fictitiously change the value of
$\alpha$.  No matter how slight the change $\Delta\alpha$ might be,
the residual change $\Delta E_r$ of the resonance energy could be
substantial, particularly in terms of the relative change $\Delta E_r
/E_r$.  The same must be true then for the cross section
$\sigma_{149}$ itself.  
\\[.5em]
\begin{figure}[htb]
\hspace*{1.5cm}
\epsfxsize=10cm
\epsffile{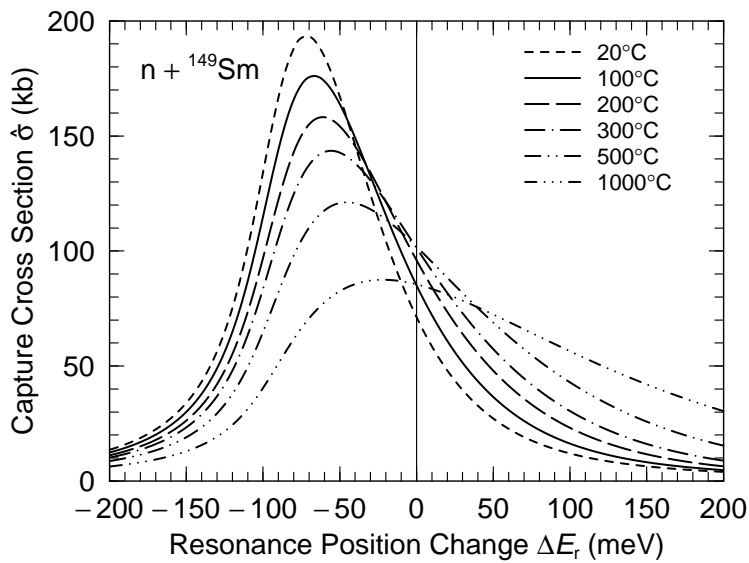}
\caption{\baselineskip= .3cm \footnotesize Effective cross section for ${\rm n}+^{149\!}{\rm Sm}\rightarrow ^{150\!}{\rm Sm} + \gamma$, for thermally
averaged  neutrons, plotted against $\Delta E_{\rm r}$.
}
\label{fig1}
\end{figure}

This is an excellent example of an enhancement mechanism, only by which
one can measure something very small.  In Fig. 1, we plotted the cross
section $\hat{\sigma}_{149}$ as a function of $\Delta E_r$, also
assuming that the neutron flux is in the thermal equilibrium of the
temperature $T$.  The hat attached above $\sigma$ implies that the Boltzmann
distribution is normalized to a particular temperature $T_0 =
20.4^{\circ}{\rm C}$, as a convention in most of the Oklo analyses,
hence the ordinary cross section $\sigma$ is given by
$\sigma(T)=\sqrt{(\pi /4)(T_0 /T)} \hat{\sigma}(T)$.  We see how
 sensitively $\hat{\sigma}$ depends on $\Delta E_r$.  If $E_r$ 2 billion years ago  were to have been smaller than the present value by as little as 10meV ($\Delta E_r = -10{\rm meV}$) and $T=20^{\circ}{\rm C}$, for example, the cross 
section would have been larger by about 10\%.

The trouble is, however, his paper in 1976 was too short to give any details on what the numerical result was like as well as of what quality his data was.  It is rather difficult to know how reliable his final result was.  We finally decided to re-do what he did previously.  We reconstructed his formulation based on his unpublished paper.$^6$

We soon realized that a major source of error comes from the following
situation.$^7$  Each of the natural reactors must have come to an end in millions of years.  Even after this termination, some amount of the isotope might have migrated from outside into the core part.  This amount has nothing to do with the nuclear interaction we are interested in.  We tried to minimize this contamination due to weathering and related phenomena as much as we could.  For this purpose we chose five samples taken from deep underground  with a great care of geologist's expertise.  In this way we obtained the data of quite good quality.

We analyzed the evolution equations for the system of $^{235}\hspace{-.0em}{\rm U}, ^{147}\hspace{-.2em}{\rm Sm}, ^{148}\hspace{-.2em}{\rm Sm}, ^{149}\hspace{-.2em}{\rm Sm}$, obtaining
\[
\hat{\sigma}_{149} = 91\pm 6 {\rm kb}.
\]
We used several ways to estimate the temperature, finding
\[
T = 200-400^{\circ}{\rm C}.
\]
Applying these ranges to Fig. 1, we found two {\em separate}
ranges:\footnote{The result for the right-branch range replaces Eq. (27) 
in reference 7, where the erroneous lower end was shown inadvertently.}
\[
\Delta E_{\rm r} =\left\{
\begin{array}{l}
 9\pm 11\;{\rm meV}, \hspace{1em}
\hspace{2em}\quad\mbox{for the right-branch range,} \nnb\\[.8em]
  -97\pm 8\;{\rm meV}, 
\hspace{2em}\quad\mbox{for the left-branch range.} \nnb
\end{array}
\right.
\]
Two ranges were obtained because a horizontal line corresponding to the observed cross section intersects the theoretical curve of the cross section for each temperature at two points, at both sides of the peak.

The right-branch range covers $\Delta E_r =0$, giving a null result
with an upper bound as usual, whereas the left-branch range indicates,
as it is, that the value of $\Delta E_r$ was {\em different} from today's value by 12 standard deviations.

We use a theoretical analysis:$^8$ 

\[
\frac{\Delta \alpha}{\alpha} = \frac{\Delta E_r}{ {\cal M}_c},
\]
where the mass scale ${\cal M}_c$ is estimated to be
\[
 \quad{\cal M}_c \approx -1.1 {\rm MeV}.
\]
This allows us to translate $\Delta E_r$ to the change of $\alpha$:
\[
 \frac{\Delta \alpha}{\alpha} =
\left\{
\begin{array}{l}
 -(0.8\pm 1.0)\times 10^{-8}, \nnb\\[.5em]
 (0.88\pm 0.07) \times 10^{-7}. \nnb
\end{array}
\right.
\]
By dividing by $-2\times 10^9{\rm y}$ we finally arrive at
\[
 \frac{\dot{ \alpha}}{\alpha} =
\left\{
\begin{array}{l}
 (0.4\pm 0.5)\times 10^{-17} {\rm y}^{-1},\nnb\\[.5em]
 -(0.44\pm 0.04) \times 10^{-16} {\rm y}^{-1}.\nnb
\end{array}
\right.
\]

As it turns out, the result from the right-branch agrees very well
with what Shlyakhter obtained in 1976.  In this sense we have confirmed his result.  The agreement to this extent, however, seems rather accidental, because it is unlikely that the data as good as ours was available at early years of Oklo investigations.

We tried hard if we could eliminate the non-null result from the
left-branch range by using other isotopes.  We had results from
$^{155}{\rm Gd}$ and $^{157}{\rm Gd}$ for which exceptionally low-lying resonances play important roles.  Unfortunately, we reached the conclusion short of complete, because better ``efficiency" of the resonances made the residual abundance even  scarcer, and left results more ``contaminated."  The null result is certainly favored but not beyond that.

We also mention, on the other hand, that the same kind of re-analysis
had been made by Damour and Dyson, who used, however, older and more
contaminated data.$^8$  As a result, they reached a much wider range of $\Delta E_r$, essentially covering our two ranges together without separation between them.

\section{Result from distant QSO's}

By a recent analysis of the spectral lines from distant QSO's, Webb
{\it et al.} obtained the first evidence for the 
changing $\alpha$.$^2$  They exploited relativistic corrections to the
spectrum having another $\alpha^2$-dependence compared with the Balmer
series spectrum.  In this way they enhanced the number of lines to be
analyzed by a factor 10, hence the statistics by 100, compared with
the measurements of alkali doublet splitting made so far.  By using the data
from 49 QSO's with the look-back time ranging from 40\% to 90\%  of
the age of the universe $\sim 1.4 \times 10^{10}{\rm y}$, they concluded
\[
\frac{\Delta\alpha}{\alpha} = (-7.5\pm 1.8)\times 10^{-6}, 
\label{bjn-8}
\]
indicating a non-zero result beyond $4.3\sigma$.

Our Oklo result is at least 2 orders smaller, but the look-back time is also smaller; $\approx 15\%$ in the above sense.  Combining them all it looks as if the time-dependence as a function of the cosmic time was nonmonotonic, or even oscillating.

\section{An accelerating universe}

Recent analyses of Type Ia supernovae finally concluded that the
universe is accelerating,$^9$ described best in terms of a small but nonzero
cosmological constant with the size as given by
\[
\Omega_\Lambda = 0.6-0.7. 
\label{bjn-10}
\]
This turns out be in agreement with other kinds of analyses, including
the age of the universe, large-scale structure, number-counting of distant galaxies and lens statistics.  We used to have the ``first face" of the
cosmological constant problem; why  it is smaller than the theoretically natural value $\sim \MP^4$ by as much as 120 orders of magnitude.  Now we confront the ``second face;" how  we can understand a nonzero $\Omega_\Lambda$ rather close to unity.  This is closely connected with the ``coincidence problem," why are we so lucky to be in this very special epoch in which $\Omega_\Lambda \sim 1$?

A possible reply to the first question is to call for the scenario of a
decaying cosmological constant; $\Lambda(t) \sim t^{-2}\sim 10^{-120}$.  This
has been implemented successfully in terms of a version of the
scalar-tensor theory.$^{10}$  According to this scenario,
today's $\Lambda$ is small only because we are old.  More
specifically, we think it 
convenient to use the reduced Planckian unit system in which $c=\hbar =\MP
(= (8\pi G)^{-1/2}) =1$.  The age of the universe $t\approx 1.3\times 10^{10}{\rm y}$ is given in units of the Planck time as $\log t \sim 60$, from
which follows $\rho_{\rm cr}\sim t^{-2}\sim 10^{-120}$, compared with
a theoretical estimate $\Lambda_{\rm th}\sim 1$, the easiest way to understand a number $\sim 10^{120}$.

We studied a prototype Brans-Dicke model with $\Lambda$ added to the
Lagrangian:
\beq
{\cal L}=\sqrt{-g}\left( \half \xi \phi^2 R - \epsilon\half g^{\mu\nu}\partial_\mu\phi  \partial_\nu\phi   -\Lambda +L_{\rm matter} \right),
\label{bjn-11}
\eeq 
with the scalar field $\phi$ and the BD parameter $\omega$, 
where we use the notations, $\epsilon = \mbox{Sign}(\omega)$ and
$|\omega | = (1/4)\xi^{-1}$.  The first term on the right-hand side is called a nonminimal coupling term which implies that we have an effective gravitational constant $G_{\rm eff}= (8\pi \xi \phi^2)^{-1}$, which is spacetime dependent, in general.

We emphasize that a conformal transformation $ds^2 \rightarrow ds^2_* = \Omega(x)^{-2}ds^2$, with a local function $\Omega(x)$, is inherently important in any theory with a nonminimal coupling term.  By a conformal transformation we move from one conformal frame to another.  A special conformal frame in which the Lagrangian is given by \reflef{bjn-11}) is called the J(ordan) frame, while the conformal transformation chosen by 
\beq
g_{\mu\nu} = \Omega^2 g_{*\mu\nu}, \quad\mbox{with}\quad \Omega^2 = \xi\phi^2,
\label{bjn-12}
\eeq
takes us to the E(instein) frame, in which the same Lagrangian \reflef{bjn-11}) is re-expressed as
\beq
{\cal L}=\sqrt{-g_*}\left( \half R_* -\half g_*^{\mu\nu}\partial_\mu\sigma  \partial_\nu\sigma   -\Lambda e^{-4\zeta\sigma} +L_{*\rm matter} \right),
\label{bjn-13}
\eeq
where the new scalar field $\sigma$ is defined by
\beq
\phi = \xi^{-1/2}e^{\zeta\sigma},\quad \zeta^{-2} = 6+\epsilon\xi^{-1},
\label{bjn-14}
\eeq
with the coupling strength $\zeta$.  Substituting this into the second 
of \reflef{bjn-12}) yields
\beq
\Omega =e^{\zeta\sigma}.
\label{bjn-14a}
\eeq

We notice that the nonminimal coupling term has now be removed and the 
$\Lambda$ term  acts as a potential, $\Lambda e^{-4\zeta\sigma}$.
The energy density of $\sigma$ allows an interpretation as an
effective cosmological constant $\Lambda_{\rm eff}$ in the E frame.  It has been argued frequently that the J frame is a physical conformal frame where we live.  Our study of the cosmology with $\Lambda$ added  reveals, however, that the E frame is to be preferred with a remedy assumed in the matter Lagrangian.$^{11}$  It also turns out that the quantum anomaly effect coming from interactions among matter fields is crucially important.

We also derive the (classical) cosmological equation in spatially flat Friedmann
universe.  The attractor solution can be obtained, which yields
\beq
\rho_\sigma =\Lambda_{\rm eff}= \half\dot{\sigma}^2 +\Lambda
e^{-4\zeta\sigma} \sim t_*^{-2}\sim \rho_{*m}, 
\label{bjn-15}
\eeq
where $t_*$ is the cosmic time in the E frame.  This equation shows 
that the effective $\Lambda$ does decay like $t_*^{-2}$, as expected.
This is the behavior sometimes called  ``scaling'' in the current
discussion of ``quintessence,$^{12}$" and is  a necessary 
condition for the solution of the coincidence problem.

We find, however, that this behavior is precisely the same way as the
ordinary matter density $\rho_{*m}$ falls off, as also shown in \reflef{bjn-15}).  This situation does not correspond to the observational finding which requires, in the present context, that $\rho_\sigma$ stays nearly constant at
least temporarily, not necessarily forever, thus {\em mimicking} a
cosmological ``constant.'' In order to meet this second face, we make
use of a feature called a ``hesitation behavior,'' which is unique to
a fast falling potential like an exponential potential.  This implies 
that there is a duration of time when the scalar field stays nearly
constant, thus is  expected to act like a cosmological constant.  We find, 
however, this ``built-in'' mechanism is short of resulting in such a
``large'' $\Omega_\Lambda \sim 0.7$.
\begin{figure}[htb]
\hspace*{3.5cm}
\epsfxsize=6cm
\epsffile{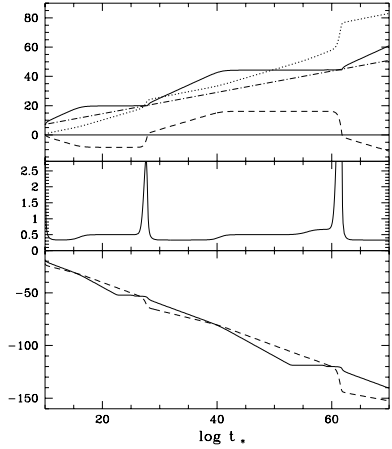}
\caption{\baselineskip= .3cm \footnotesize An example of the solution of the classical equation for the potential $V(\sigma, \chi)$.  Upper diagram: $b=\ln a_*$
(dotted), $\sigma$ (solid)  and $2\chi$ (dashed) are plotted against $\log t_*$.  The present epoch corresponds to $\log t_* =60.1-60.2$, while the primordial nucleosynthesis must have taken place at $\log t_* \sim 45$.   The parameters are $\Lambda =1, \zeta = 1.5823, m= 4.75, 
\gamma = 0.8, \kappa = 10$.  The initial values at $t_1 = 10^{10}$
are $\sigma _1=6.7544, \sigma'_1 =0$ (a prime implies differentiation
with respect to $\tau=\ln t_*$), $\chi_1 = 0.21, \chi'_1 = -0.005,
\rho_{1 \rm rad}= 3.7352 \times 10^{-23}, \rho_{1 \rm dust}=4.0 \times
10^{-45}$. The dashed-dotted straight line represents the asymptote of $\sigma$ given by $\tau /(2\zeta)$.  Notice long plateaus of $\sigma$ and
$\chi$, and their rapid changes  during relatively ``short''
periods.   Middle diagram: $\alpha_* =b' = t_* H_*$ for an effective
exponent in the local power-law expansion $a_*\sim t_*^{\alpha_*}$ of
the universe.   Notable leveling-offs can be seen at 0.333, 0.5 and 0.667
 corresponding to the epochs dominated by  the kinetic terms of the
scalar fields, the radiation matter and the dust matter, respectively.
Lower diagram: $ \log\rho_s$ (solid), the total energy density of the
$\sigma-\chi$ system, and $\log\rho_{*m}$ (dashed), the matter energy density.
  Notice an interlacing pattern of $\rho_s$ and $\rho_{*m}$, still obeying 
$\sim t_*^{-2}$ as an overall behavior.  Nearly flat plateaus of
$\rho_s$ precede before $\rho_s$ overtakes $\rho_{*m}$, hence with
$\Omega_{\Lambda}$ passing through 0.5. }
\label{fig2}
\end{figure}

We searched for a remedy on a try-and-error-basis, coming across
finally a ``two-scalar model.''  We introduced another scalar field,
called $\chi$, together with an interaction potential
 in the E frame:
\[
V(\sigma, \chi)=e^{-4\zeta\sigma}\tilde{V}(\sigma, \chi), 
\]
where 
\[
\tilde{V}= \Lambda+\half m^2\chi^2\left[ 1+\gamma\sin(\kappa\sigma ) \right],
\]
with $m, \gamma,$ and $\kappa$ constants chosen not to be very much
different from unity in the reduced Planckian unit system.  Note that
this potential goes into the exponential potential as before in the
limit $\chi =0$.  This potential might look complicated and  artificial, but turns out to  result in many desired features, as one will find in a typical
example of the solution illustrated in Fig. \ref{fig2}.   See Ref. 13 for more details.

\begin{figure}[t]
\parbox[t]{6cm}
{
\epsfxsize=5.5cm
\epsffile{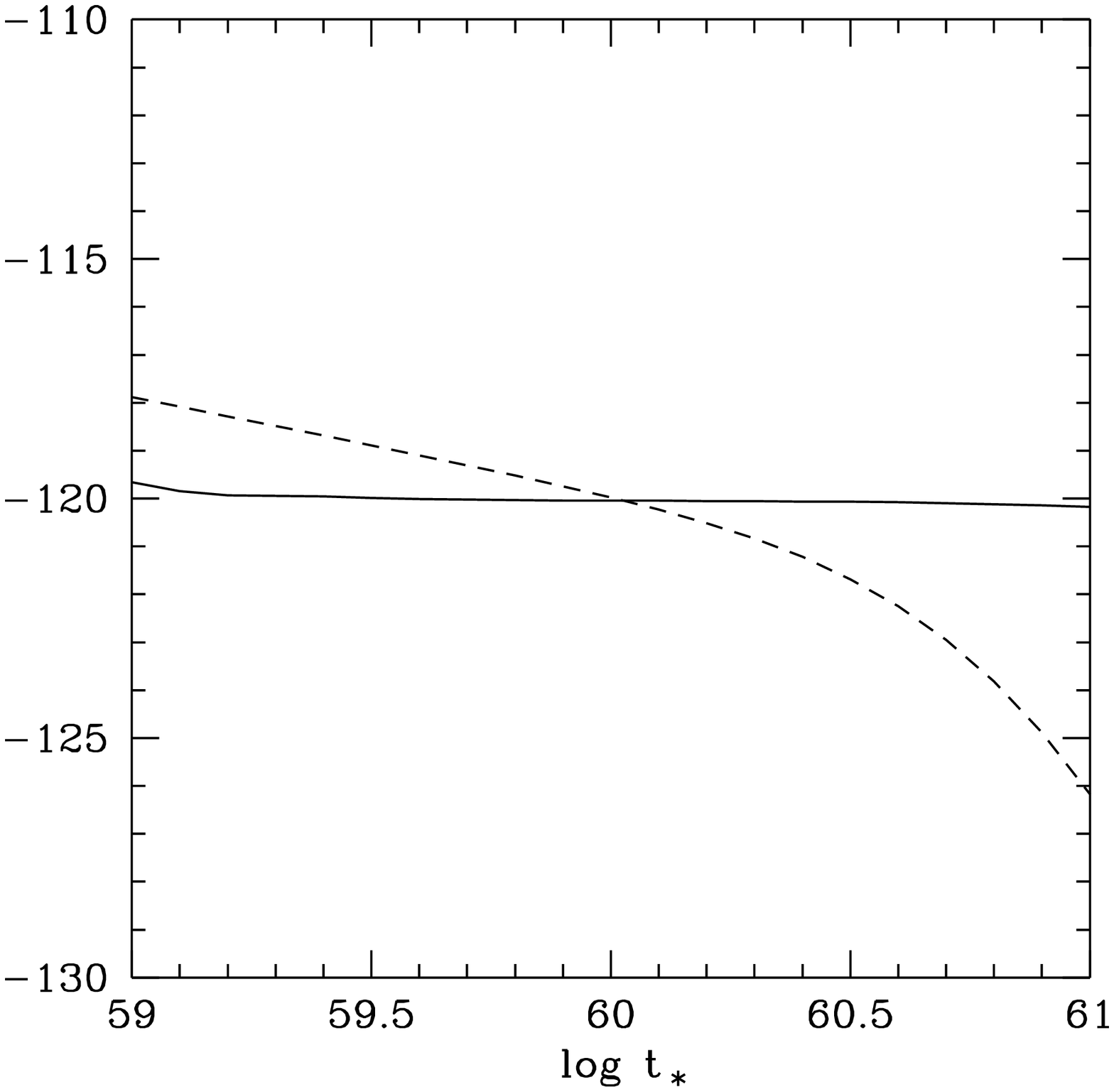}
\caption{\baselineskip= .3cm \footnotesize Magnified view of $\log\rho_s$ (solid) and $\log\rho_{*m}$ (dashed) in the lower diagram
of Fig. \protect\ref{fig2} around the present epoch.  Note that
$\rho_s$ is very flat in this diagram extending back to the past
of $z=5.2-6.9$ for the assumed age $(1.1-1.4)\times 10^{10}{\rm y}$.
}
\label{fig3}
}
\mbox{\hspace{1cm}}
\parbox[t]{6cm}
{
\epsfxsize=5.5cm
\epsffile{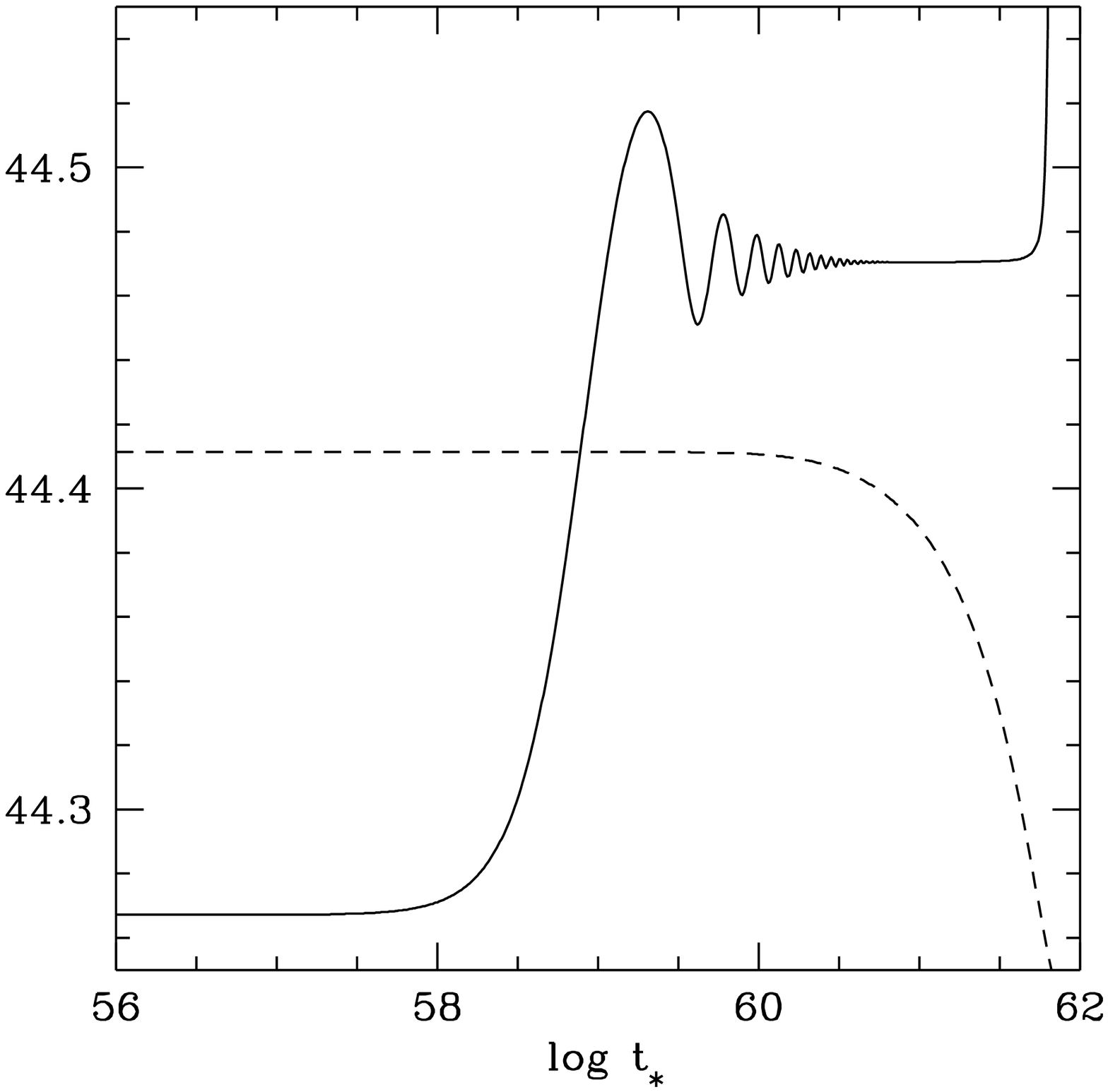}
\caption{\baselineskip= .3cm \footnotesize Magnified view of $\sigma$ (solid) and $0.02 \Phi+44.25$ (dashed) in the upper diagram of
Fig. \protect\ref{fig2}.  Note that the vertical scale has been expanded by approximately 330 times as
large compared with Fig. \protect\ref{fig2}.  
With the time variable $\tau = \ln t_*$, the potential $V$ grows as
the multiplying factor $t_*^2=e^{2\tau}$.  The potential wall for
$\sigma$ becomes increasingly steeper, thus confining $\sigma$ further
to the bottom of the potential, as noticed by an oscillatory behavior of $\sigma$ with ever increasing frequency measured in $\tau$.  The growing $V$ causes $\chi$ eventually to fall downward, resulting in the collapse of the confining potential wall.  The stored energy is then released to unleash $\sigma$.
}
\label{fig4}
}
\end{figure}

We fine-tuned parameters and initial values  moderately.  Around the
present epoch, $\log t_{*0}\sim 60$, the density $\rho_s =
\Lambda_{\rm eff}$ for 
the scalar-field system surpassed the ordinary density $\rho_{*m}$ as shown in Fig. \ref{fig3}, giving $\Omega_\Lambda = 0.75$ with the Hubble parameter $ H_0= 72.9
{\rm km}/{\rm sec}/{\rm Mpc}$.  The coincidence problem is now less
severe  because  a ``mini-inflation'' is
one of the {\em repeated} events, instead of a once-and-for-all occurrence in 
the whole history of the universe.

Figs. 3,4 are magnified view around the present epoch. In particular, 
Fig. 4 shows an oscillatory behavior of $\sigma$, which might be
related to the time variation of $\alpha$,  as will be discussed.

\section{Time-dependent fine-structure constant?}

We consider QED as matter part of the Lagrangian \reflef{bjn-11}) in
the J frame, by introducing a charged spinor field $\psi$ and the
Maxwell field:
\beq
{\cal L}_{\rm matter}=\sqrt{-g} \left( -\psibar \left( \Dslash 
+ie A \hspace{-.45em}/ + m \right) \psi 
-\frac{1}{4}FF \right),
\label{bjn-20}
\eeq
where the charge $e$ is assumed truly constant in this J frame.

We apply the conformal transformation.  In the E frame we include the
loop integrals which are divergent, hence requiring {\em regularization}.
For this purpose we assume $D$-dimensional spacetime.  We stay in $D$
dimensions until we go to the limit $D\rightarrow 4$ at the end of the 
calculation.

To be noticed is the fact that the Maxwell theory is conformally
invariant in 4 dimensions.  It then follows that the electromagnetic
field $A_\mu$ in $D$ dimensions is transformed into $A_{*\mu}$ defined by
\beq
A_{*\mu} = \Omega^{2-d}A_{\mu},
\label{bjn-21}
\eeq
where $d=D/2$.  It also follows
\beq
e_* = \Omega^{d-2}e = e e^{(d-2)\zeta\sigma} \approx e\left[ 1+(d-2)\zeta\sigma  \right]. 
\label{bjn-22}
\eeq
It appears that $e$ remains unchanged for $d\rightarrow 2$.
According to the spirit of dimensional regularization, however, we
leave $d-2$  
as it is at the moment.  \reflef{bjn-22}) implies that a change of the 
field $\sigma$ induces a classical change of the electric charge $e_*$:
\beq
\Delta e_* = \zeta (d-2) e_* \Delta\sigma.
\label{bjn-23}
\eeq

We consider the electromagnetic vertex of $\psi$, inserting a $\psi$
loop in the photon self-energy part, which develops a pole $(2-d)^{-1}$.
This cancels the zero $(d-2)$ in \reflef{bjn-23}), leaving a finite
contribution, in the same manner as we obtain quantum {\em anomalies}.  In this way we arrive at
\beq
 \frac{\Delta\alpha_*}{\alpha_*} = \frac{\alpha_*}{2\pi}\Delta\sigma  \approx 1.2\times 10^{-3}\Delta\sigma.
\label{bjn-24}
\eeq

From the example shown in Fig. \ref{fig4}, we find $\Delta\sigma
\approx 5\times 10^{-2}$ during the time span $10^{59} - 10^{60}$
corresponding to the look-back time of 90\% of the age of the universe.  Using this on the
right-hand side of \reflef{bjn-24}), we finally obtain
\beq
\frac{\Delta\alpha_*}{\alpha_*} \approx 0.6\times 10^{-4},
\label{bjn-25}
\eeq
which is found to be  an order of magnitude too large compared with 
the observation in Ref. 2.

We need to select other solutions by changing the initial values
slightly, without changing the result for $\Omega_\Lambda$ and $H_0$
in any appreciable amount.  Also we have to improve the theoretical
calculation with respect to other particles inside the loop.

At this moment, we only say that comparison with the result on
$\Delta\alpha/\alpha$ constrains the cosmological solution much
further beyond fitting the accelerating universe alone.  Details of the
analysis will be published elsewhere.$^{14}$\footnote{After having
submitted the paper, we came, by a careful analysis of the conformal 
frame taking the quantum nature of particle masses into account, to
find$^{14}$ that  our analysis leads eventually to a time-dependent $G$ with  $| \dot{G}/G| \lsim 10^{-14} {\rm
y}^{-1}$, which might be tested in the future by the measurements
 as proposed in Refs. 15.}

\section{Concluding remarks}

Based on an assumption in the scalar-tensor theory, we have come to providing an explicit implementation of the idea that time-dependence of the fine-structure constant ensues from the evolution of the scalar field, which is responsible for the accelerating universe.  Still even at this preliminary stage, the order-of-magnitude agreement between the sizes of the two phenomena seems remarkable.

As another outcome unique to the scalar-tensor theory, we point out that the proposed remedy to the prototype Brans-Dicke model entails the occurrence of non-Newtonian gravity featuring a finite force-range and the violation of Weak Equivalence Principle probably within the observed constraints obtained so far.$^{16}$

\end{document}